\documentclass[a4paper,fleqn,usenatbib]{mnras}

\usepackage[T1]{fontenc}
\usepackage{ae,aecompl}

\usepackage{graphicx}	
\usepackage{amsmath}	
\usepackage{amssymb}	
\usepackage{float}
\usepackage{natbib}
\usepackage{morefloats}
\usepackage{dcolumn}
\usepackage{amsmath}
\usepackage{latexsym}
\usepackage{longtable}
\usepackage{lipsum} 

\begin{document}
\title[Partial eruption of a bifurcated solar filament]
{On the partial eruption of a bifurcated solar filament structure}
\author[Monga A. et al.]
{Aabha Monga\thanks{e-mail: aabha@aries.res.in}$^{1,2}$, Rahul Sharma\thanks{rahul.sharma@uah.es}$^{3}$, Jiajia Liu$^{4, 5}$, Consuelo Cid$^{3}$, Wahab Uddin$^{1}$, \newauthor{Ramesh Chandra$^{2}$}, Robertus Erd\'elyi$^{5, 6, 7}$ \\
$^{1}$Aryabhatta Research Institute of Observational Sciences 
(ARIES), Manora Peak, Nainital-263001, India\\
$^{2}$Department of Physics, DSB Campus, Kumaun University, Nainital-263001, India\\
$^{3}$Space Weather Research Group, Departamento de F\'isica y Matem\'aticas, Universidad de Alcal\'a, A-2 km 33,600,\\ 28871 Alcal\'a de Henares, Madrid, Spain\\
$^{4}$Astrophysics Research Centre, School of Mathematics and Physics, Queen's University, Belfast, BT7 1NN, Northern Ireland, UK \\
$^{5}$Solar Physics \& Space Plasma Research Centre, School of Mathematics and Statistics, University of Sheffield, Sheffield S3 7RH, UK\\
$^{6}$Department of Astronomy, E\"otv\"os Lor\'and University, P\'azm\'any P\'eter s\'et\'any 1/A, H-1117 Budapest, Hungary\\
$^{7}$Gyula Bay Zolt\'an Solar Observatory (GSO), Hungarian Solar Physics Foundation (HSPF), Pet\H{o}fi t\'er 3., Gyula, H-5700, Hungary}

\date{Accepted XXX. Received YYY; in original form ZZZ}

\pubyear{2020}

\label{firstpage}
\pagerange{\pageref{firstpage}--\pageref{lastpage}}
\maketitle

\begin{abstract}
The partial eruption of a filament channel with bifurcated substructures is investigated using datasets obtained from both ground-based and space-borne facilities. Small-scale flux reconnection/cancellation events in the region triggered the pile-up of ambient magnetic field, observed as bright EUV loops in close proximity of the filament channel. This led to the formation of a V-shaped cusp structure at the site of interaction between the coalesced EUV loops and the filament channel, with the presence of distinct plasmoid structures and associated bidirectional flows. Analysis of imaging data from \textit{SDO}/AIA further suggests the vertical split of the filament structure into two substructures. The perturbed upper branch of the filament structure rose up and erupted with the onset of an energetic \textit{GOES} M1.4 flare at 04:30 UT on January 28, 2015. The estimated twist number and squashing factor obtained from nonlinear force free-field extrapolation of the magnetic field data support the vertical split in filament structure with high twist in upper substructure. The loss in equilibrium of the upper branch due to torus instability, implying this as a potential triggering mechanism of the observed partial eruption.
\end{abstract}




\section{Introduction}
Solar filaments are high-density (10$^{9}$ - 10$^{11}$ cm$^{-3}$), low-temperature ($\sim$10$^{4}$ K) structures that remain suspended by magnetic fields in the solar corona. These features when observed on-disk, are found along the polarity reversal lines between the regions of oppositely directed photospheric magnetic fields \citep{Martin1998}. Topologically, filaments are modeled as helical magnetic flux ropes (MFRs) in which the plasma is trapped along the twisted field lines, wrapped around a guiding axis \citep{Kuperus1974, van1989}, and/or as sheared-arcades with long sheet of dense material held in dipped magnetic fields \citep{Mackay2010, Gibson2018}. These structures are often associated with eruptive events that can potentially disrupt the near-Earth space environment, with around 70\% of coronal mass ejections (CMEs) linked to filaments \citep{Munro1979, Zhou2003}. Though, till today, multi-wavelength observations from both ground-based and space-borne instruments, along with theoretical and numerical studies \citep{Green2018} have provided an insight into the evolution, topology and dynamics of solar filament eruption, however, the inter-relationship of the causative factors still remains unknown. 

In general, filament eruptions are broadly classified into three main categories, viz, full \citep{Plunkett2000, Rust2003}, failed- \citep{Moore2001, Alexander2006, Kumar2011a, Zheng2015} and partial-eruptions \citep{Gilbert2000, Liu2007}. A possible explanation of the above was given by \citet{Gilbert2001, Gilbert2007}, where they suggested that these different eruption scenarios could be due to the location of the magnetic reconnection site around the MFR structure. If the reconnection site is below the MFR, then it will result in a full-eruption with the filament escaping the solar-disk as a classical three-part structure \citep{Cremades2004, Zurbuchen2006}. In case, the reconnection neither lifts the filament mass nor the MFR, which is trapped due to the overlying magnetic arcade, the eruption is categorized as a confined/failed-eruption. However, it must be noted that other factors can also contribute to the failed-eruption scenario, that include e.g. decrease in magnetic field strength with height \citep{Torokkliem2005}, and/or insufficient energy release during eruptions \citep{Shen2011}.

If the reconnection site is located either above or within the MFR structure, it can result into a partial eruption. In this case, a filament breaks (at least) into two parts with the upper flux escaping the solar atmosphere as a part of the CME, while an other part remains more closely bounded to the Sun. The persistent filament fragment tends to reform by accumulation of plasma and reappears in the observations. This mechanism was also investigated through numerical simulations by \citet{Gibson2006a, Gibson2006, Gibson2008}, where they suggested that if the low-lying MFR connects to the photospheric polarity inversion line, the separatrix surface (also known as bald-patch), prevents the lower part of the MFR from erupting, resulting into splitting which is followed by the partial eruption of the filament structure. Furthermore, \citet{Birn2006} suggested that even in the absence of a blad-patch topology, a kink-unstable MFR can break into escaping and leaving fragments due to the high number of twists in the structure.

Observations revealed that around one-third of the filament eruptions were partial, as indicated by \citet{Gilbert2000}, where they found that out of 54 H$\alpha$ filament eruptions 18 were categorized as partial eruptions. This was further confirmed by similar cases reported in subsequent studies \citep[e.g.,][]{Pevtsov2002, Tripathi2009}, with part of the structure either fall back or remained on the Sun. Moreover, the observed mechanism of partial eruptions could possibly be sub-divided into two subtypes: vertical and horizontal splitting. Vertical splitting happens if the reconnection site is located within/above the filament MFR, or due to the instability between two vertically arranged MFRs with similar chirality and current direction, forming a double-decker configuration \citep{Kliem2014}. Horizontal splitting of the filament structure is believed to happen primarily due to magneto-hydrodynamic (MHD) instabilities, rather than magnetic reconnection. Moreover, the horizontal splitting is often observed with asymmetric filament eruptions \citep{Contarino2003, Tripathi2006}, and/or complex photospheric magnetic configurations \citep{Guo2010}. In case of asymmetric eruptions, the filament feature undergoes whipping/zipping motions along the spine. These motions, due to the interaction with the ambient field and/or any overlying quasi-separatrix layer, can detach a part of the filament structure which might appear akin to horizontal split \citep[see,][and references therin.]{Liu2018}

In recent years, observations of the splitting behavior in MFR prior to partial eruptions of filaments received much attention. \citet{RuiLiu2012} reported the eruption in a vertically split filament structure, stable in a `double-decker' configuration, hours before the eruption of the upper branch. Similar ``double-decker'' equilibria were also reported in many other studies \citep[e.g.,][]{Su2011,Cheng2014,Zhu2014a,Zhu2015,Liu2016,Cheng2018,Dhakal2018,Hou2018}. \citet{Mashnich2014} analyzed a combination of data from both space-borne and ground-based observations and reported horizontal breaking of the filament feature. \citet{Chintzoglou2015} showed that horizontal split could arise in under-developed filament channels for region with complex topology consisting of both MFR and arcade structures. Recently, \citet{Prasad2017} reported splitting in an asymmetric filament eruption in a complex active region and compared the results with numerical simulations for emerging flux. Also, \citet{Dacie2018} showed the origin of multiple CMEs from a horizontally split filament structure using observations and 3D MHD simulations. They highlighted the role of emerging flux in filament splitting and eruption of multiple CME from the same structure.  

Despite of much attention given to the magnetic flux splitting in filament structures, both to vertical and horizontal domains, the mechanisms responsible for initiation of this behavior and later eruption, remains elusive. In the present paper, we investigate the partial eruption of a bifurcated filament structure observed in a complex magnetic topology region of AR12268. The feature split into two visible fragments, one with a later eruption of the upper branch associated with an M1.4 class flare. Before the onset of the flare, the region had multiple small-scale reconnection events observed as brightenings in multiple EUV passbands. Here, the possible role of these small-scale brightening, as precursor events, leading to the filament splitting is examined. Observables, such as, formation of a V-shaped cusp, bright plasmoid(s), bidirectional (out)flows and associated MHD instabilities are investigated to identify the factors leading to the partial eruption of the observed filament structure.

\begin{figure*}
\begin{center}
\includegraphics[scale=0.32]{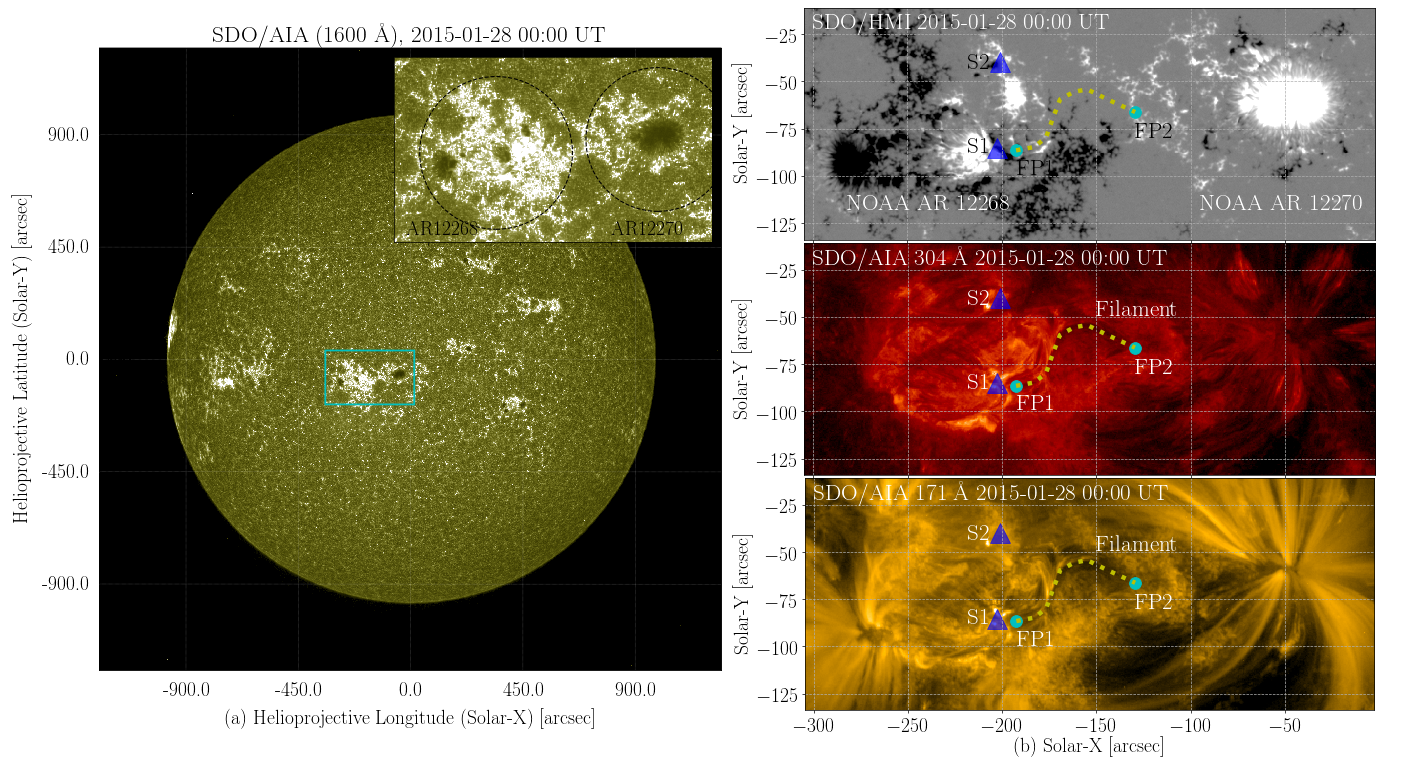}
\caption{Overview of the two active regions (AR12268 and AR12270), observed at multiple heights in the solar atmosphere. Left image (a) shows the observation of the region (highlighted in box and zoomed inset), located near the disk-center with two active regions, on January 28, 2015, in photospheric \textit{SDO}/AIA 1600 $\textup \AA$ wavelength. Complex magnetic topology of the region is shown (b): top-bottom, \textit{SDO}/HMI magnetogram (top), along with observations in chromospheric (\textit{SDO}/AIA 304 $\textup \AA$: middle) and coronal (\textit{SDO}/AIA 171 $\textup \AA$: bottom) wavelengths. The filament structure is highlighted by the dotted line, along with footpoints (FP1 and FP2). The sites for magnetic flux cancellation, prior to the partial eruption of filament, are marked by triangles (S1, S2).}
\label{fig:1}
\end{center}
\end{figure*}

\begin{figure*}
\begin{center}
\includegraphics[scale=0.275]{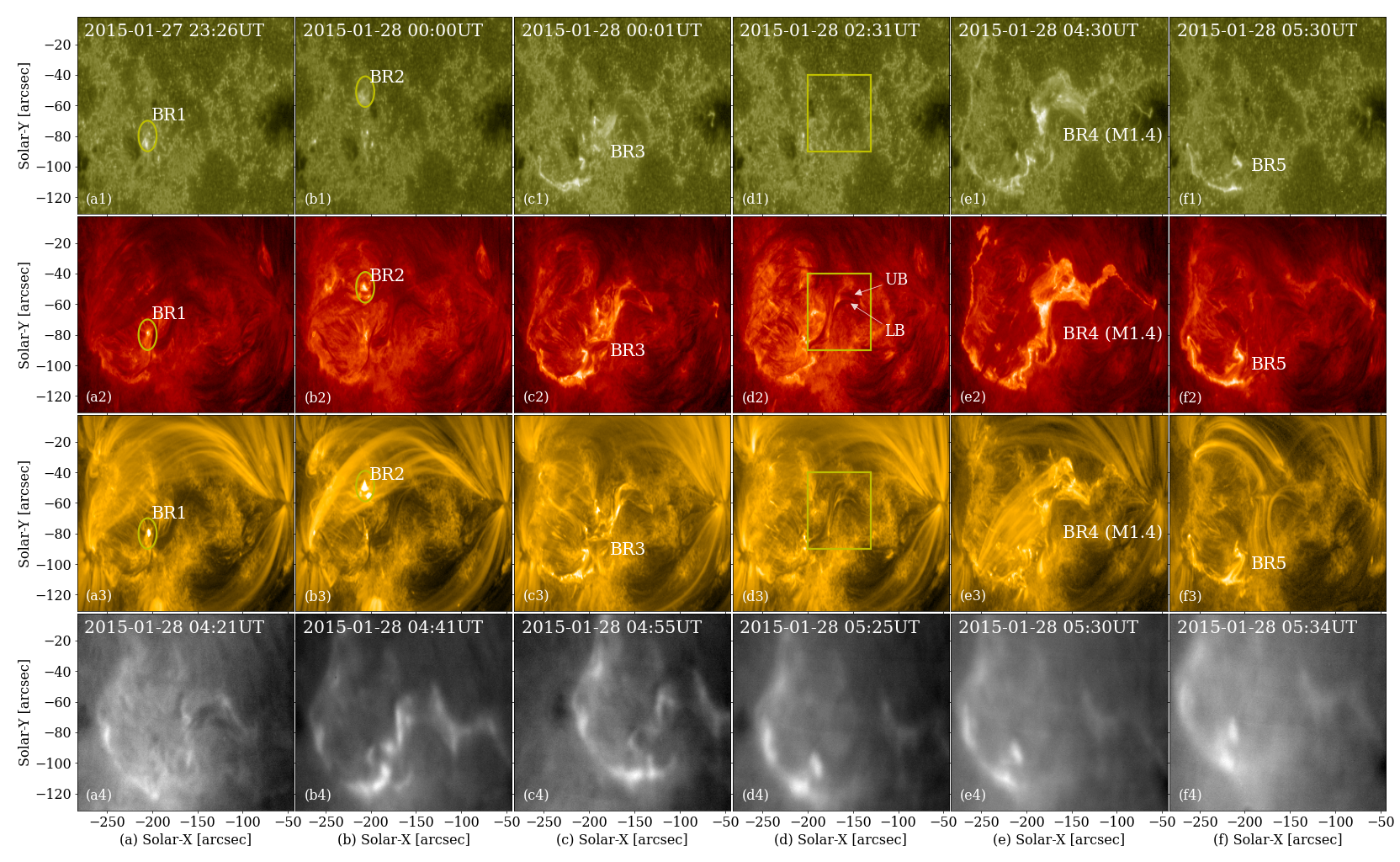}
\caption{Temporal evolution of AR12268 with the filament MFR in different layers of the solar atmosphere. Panels (1-3, top to bottom) show the active region in photospheric (\textit{SDO}/AIA, 1600 $\textup\AA$), chromospheric (\textit{SDO}/AIA, 304 $\textup\AA$) and coronal (\textit{SDO}/AIA, 171 $\textup\AA$) wavelengths, while the temporal evolution (a -f, left to right), highlighting the subsequent brightening (BR1 - BR5), along with bifurcation of the filament structure (d2). The branched filament is highlighted in yellow box with substructures marked as UB (upper branch) and LB (lower branch). The M1.4 flare (BR4) is shown in \textit{SDO}/AIA passbands (panels (c1) - (c3)), along with observations from ARIES-H$\alpha$ telescope (panels (a4) - (f4)), suggesting the partial eruption of the filament MFR. An animation of this figure is available online.}
\label{fig:2}
\end{center}
\end{figure*}

\section{Data and Analysis Methods}
The data used for the analysis here are primarily from the Atmospheric Imaging Assembly \citep[AIA:][]{Lemen2012} onboard the Solar Dynamics Observatory \citep[\textit{SDO}:][]{Pesnell2012} satellite that acquires the Sun's full-disk images to 1.3 $\textup{R}_\odot $ with temporal and spatial resolution of 12 sec and 0.6$\arcsec$, respectively. Evolution of the filament MFR structure, with chromospheric and coronal passbands is observed in 304 $\textup \AA $ (\ion{He}{II}, T $\sim$0.05 MK) and 171 $\textup \AA $ (\ion{Fe}{IX}, T $\sim$0.7 MK) channels. Vector magnetograms in \ion{Fe}{I} (6173 $\textup \AA$) wavelength of the Helioseismic and Magnetic Imager \citep[HMI;][]{Schou2012a}, were used to analyze the magnetic field topology of the filament structure and the active region. The vector magnetograms were processed and released as part of the Spaceweather HMI Active Region Patch \citep[SHARP;][]{Bobra2014}, where the magnetic field estimates were derived assuming a Milne-Eddington atmosphere, and remapped to a Lambert Cylindrical Equal-Area (CEA) projection. 

The filament structure along with the evolution of an M1.4 class flare was also observed in the H$\alpha$ (6563 $\textup \AA $) wavelength, using a 15-cm Coud\'e solar telescope equipped with an H$\alpha$ filter and 1k x 1k CCD camera system at Aryabhatta Research Institute of observational sciencES (ARIES), Nainital, India. The images obtained by the ARIES telescope were magnified twice of their original size with the help of a Barlow lens. The final science-grade data, processed using SolarSoft routines, had the spatial- and temporal-resolution of 1.0$\arcsec$ and 2 sec, respectively. Further, soft x-ray emissions from Geostationary Operational Environmental Satellite (\textit{GOES}) and imaging data from the Large Angle Spectrometric Coronagraph \citep[LASCO:][]{Brueckner1995}, onboard the Solar and Heliospheric Observatory (\textit{SoHO}), were also used to estimate the energetics of the flare and near-Sun characteristics of the associated CMEs.

The kinematics of the flux structures in the plane-of-sky (POS) were investigated using time-distance (TD) plots, created by applying cross-cuts on AIA intensity images. Thermodynamic evolution of the features is studied using differential emission measure (DEM) estimations. DEMs are generated using co-temporal observations of six \textit{SDO}/AIA channels (94, 131, 171, 193, 211 and 335 $\textup \AA$) sampling coronal temperatures.   The estimated DEM can be given as,

\begin{align}
I_{i} = \int {R_{i}(T) \times \textup{DEM} (T) \ \ dT,}
\end{align}
where, I$_{i}$ is the intensity magnitude for the \textit{SDO}/AIA channel (`i'). R$_{i}$(T) is the corresponding temperature response function, while DEM(T) is the differential emission for the coronal plasma, estimated using the xrt\_dem\_iterative2.pro routine given in the SolarSoftware package. Limitations and possible error in the DEM estimates are discussed in \citet{Hannah2012}. Moreover, the emission measure for a temperature range (T$_{min}$, T$_{max}$) is evaluated as,

\begin{align}
EM  = \int_{T_{min}}^{T_{max}} \textup{DEM}(T) \ \ dT.
\end{align}

Evolution of the magnetic topology of the filament MFR is investigated using nonlinear force-free field (NLFFF) extrapolations \citep{Wiegelmann2004} that is better suited to model the current carrying structures in the low-$\beta$ coronal environment \citep{Wiegelmann2012}. Preprocessed vector magnetograms were used, as photospheric boundary conditions for the extrapolation of the magnetic field lines in the corona. Furthermore, to identify the possible role of any MHD instability in association with the flare, magnetic twist (T$_{W}$) and squashing factor ($Q$) of the field lines were also estimated, using the method described by \citet{Liu2016}. The twist number can be given as;
\begin{align}
T_{W}& = \int_{L}\frac{\mu_{0}J_{\parallel}}{4\pi B}dl = \int_{L}\frac{\nabla\times B\cdot B}{4\pi B^{2}}dl \\ 
&= \frac{1}{4\pi}\int_{L}\alpha \; dl,
\end{align}

assuming, $ \nabla\times B = \alpha B$, where, $\alpha$ is the force-free parameter, and $\nabla\times B \cdot B/4\pi B^{2}$, is taken as the local twist density. Variation in the horizontal component of the magnetic field ($B_{h}$), represented by the decay index ($n$), given as $n = -d \ln |B_{h}| / d \ln|H|$ \citep{Kliem2006}, is also computed to approximate the height (H) over the photosphere for the onset of any instability. The decay index is estimated for the area comprising both active regions, as shown in Figure \ref{fig:1}(b).

\begin{figure*}
\begin{center}
\includegraphics[scale=0.28]{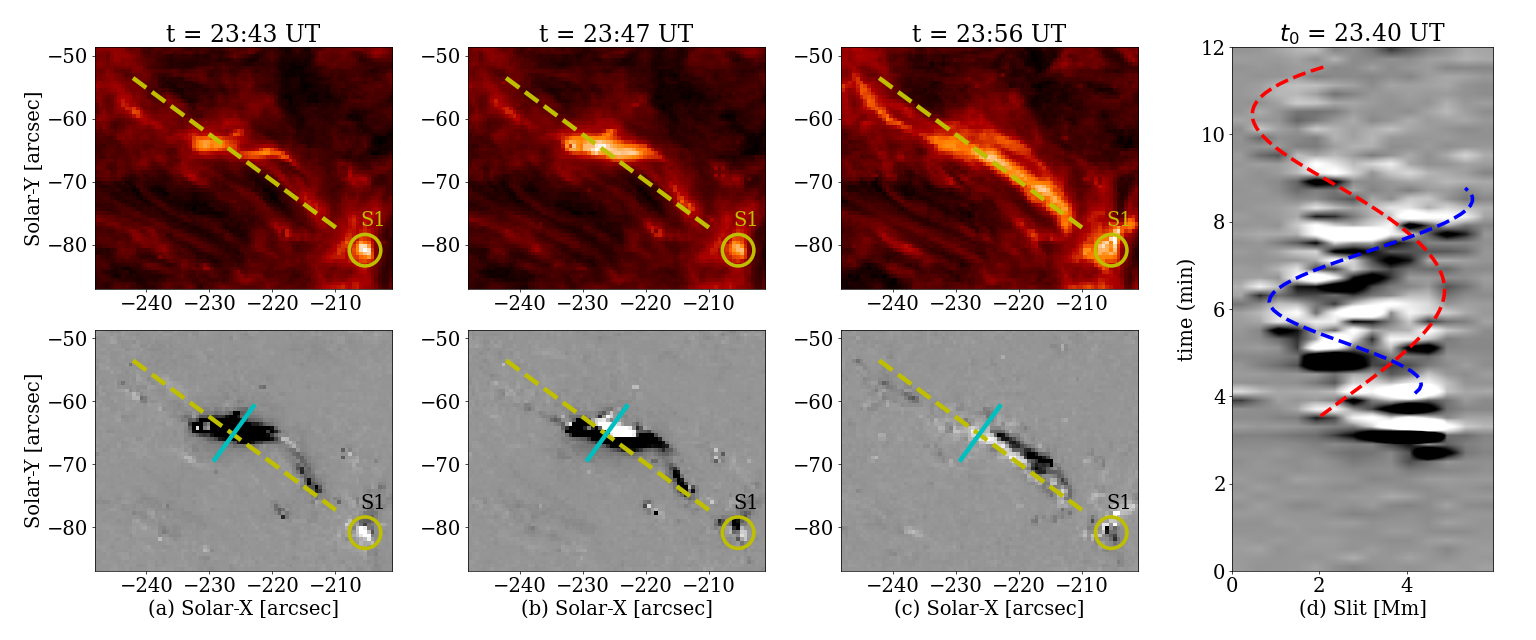}
\caption{Sequence of jet-like structure emanating out from the site `S1', associated with brightening (BR1) at $\sim$23:26 UT on 27 January. Top panels (a) - (c) show corresponding running-difference images of the feature, along with \textit{SDO}/AIA 304 $\textup\AA$ intensity images with the jet axis marked as dashed-line. The location of an artificial slit used to generate the time-distance plot is marked as a line (cyan) perpendicular to the jet axis. Sinusoidal tracks highlighted by dashed lines (red, blue) on time-distance plot (d) indicates possible mass motion along the curved path and/or swinging of the jet feature. An animation of this figure is available online.}
\label{fig:3}
\end{center}
\end{figure*}

\begin{figure}
\begin{center}
\includegraphics[scale=0.36]{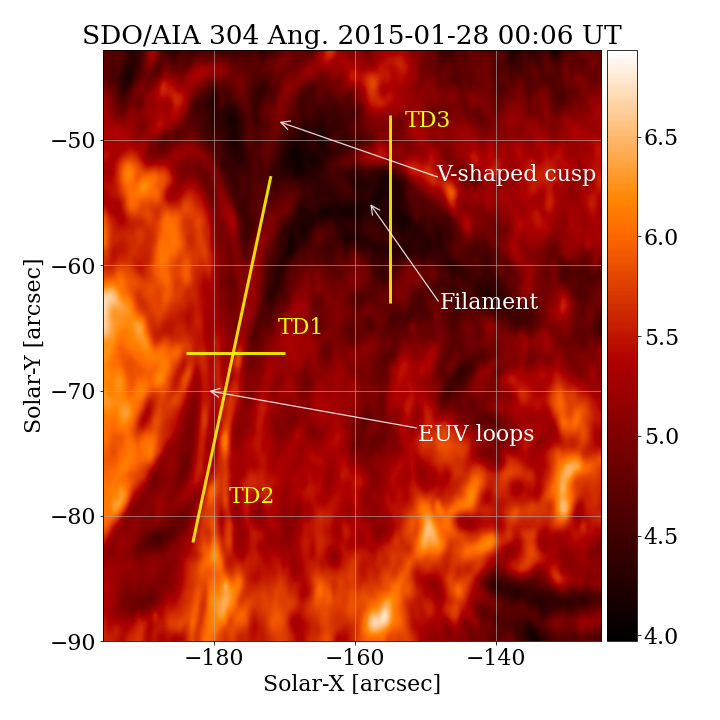}
\caption{\textit{SDO}/AIA 304 $\textup \AA$ intensity image show the V-shaped cusp morphology at the site of interaction in between the coalesced EUV loops and the filament structure on January 28, 2015, at 00:06 UT. Locations of the artificial slits used to investigate the kinematics associated with EUV loop propagation towards the filament channel (TD1), bidirectional flows due to flux interactions (TD2), and filament bifurcation (TD3) are also marked with yellow lines.}
\label{fig:4}
\end{center}
\end{figure}

\begin{figure*}
\begin{center}
\includegraphics[scale=0.38]{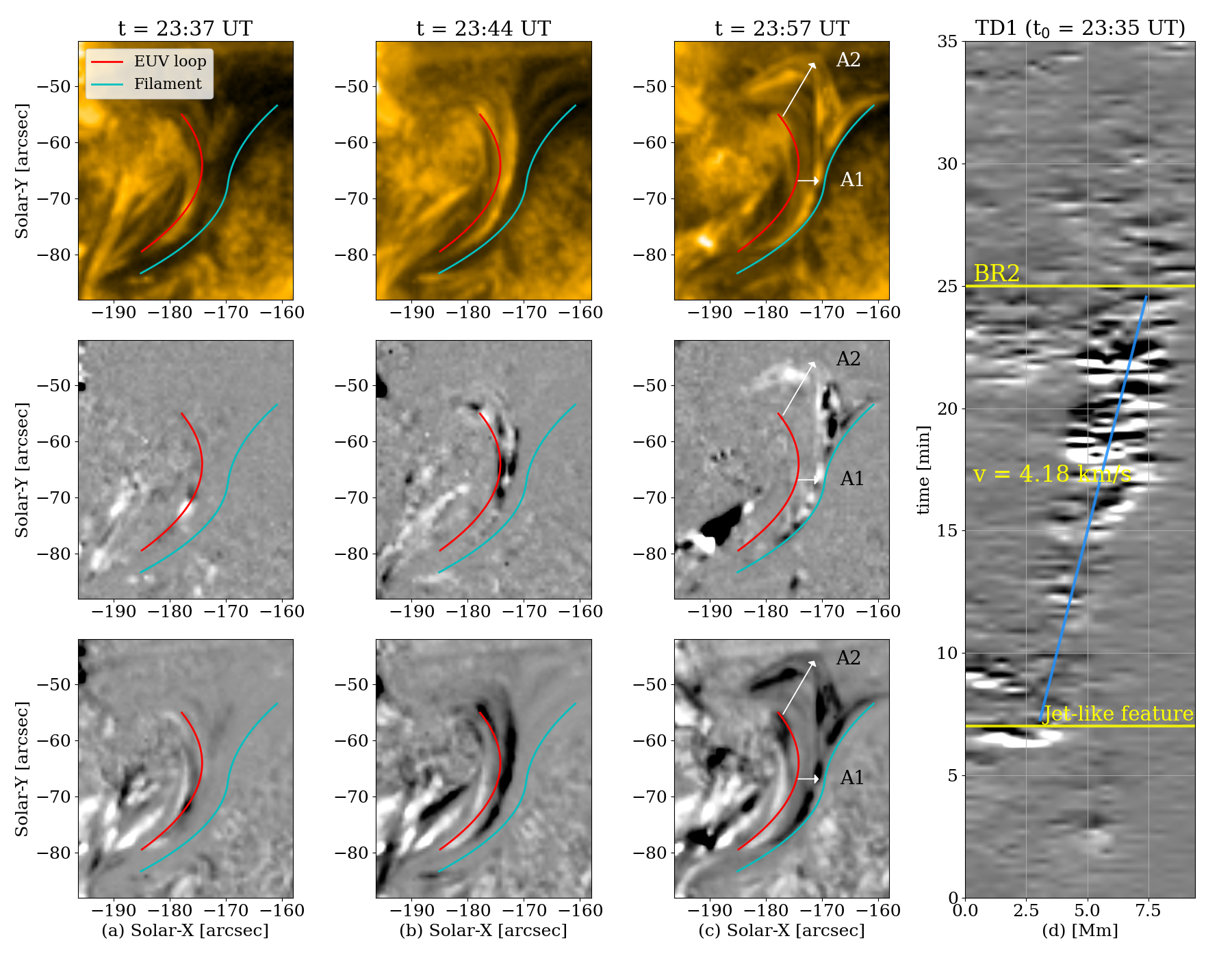}
\caption{Panels (a)-(c) show the temporal evolution of the EUV loops towards the filament MFR in intensity (top), running-difference (middle) and base-difference (bottom) images at three time-steps. Initial positions of the loop structure and the filament channel at 23:35 UT are marked with red and cyan colored lines respectively. Arrows (A1 and A2) in panels (c) show the comparative displacement of the loop at two different locations w.r.t. the filament channel during 23:35-23:57 UT on 27 January, 2015. Time-distance plot (d) suggests the migration of EUV loops towards the filament channel (marked by a line), during the interval 23:35 - 00:10 UT, with an average velocity of 4.18 km/s. The epoch-time for the flux motion coincides with the observation of jet-like feature (Fig.\ref{fig:3}), and is marked by a yellow line at 23:42 UT, along with brightening (BR2) at 00:00 UT. An animation of this figure is available online.}
\label{fig:5}
\end{center}
\end{figure*}

\begin{figure*}
\begin{center}
\includegraphics[scale=0.28]{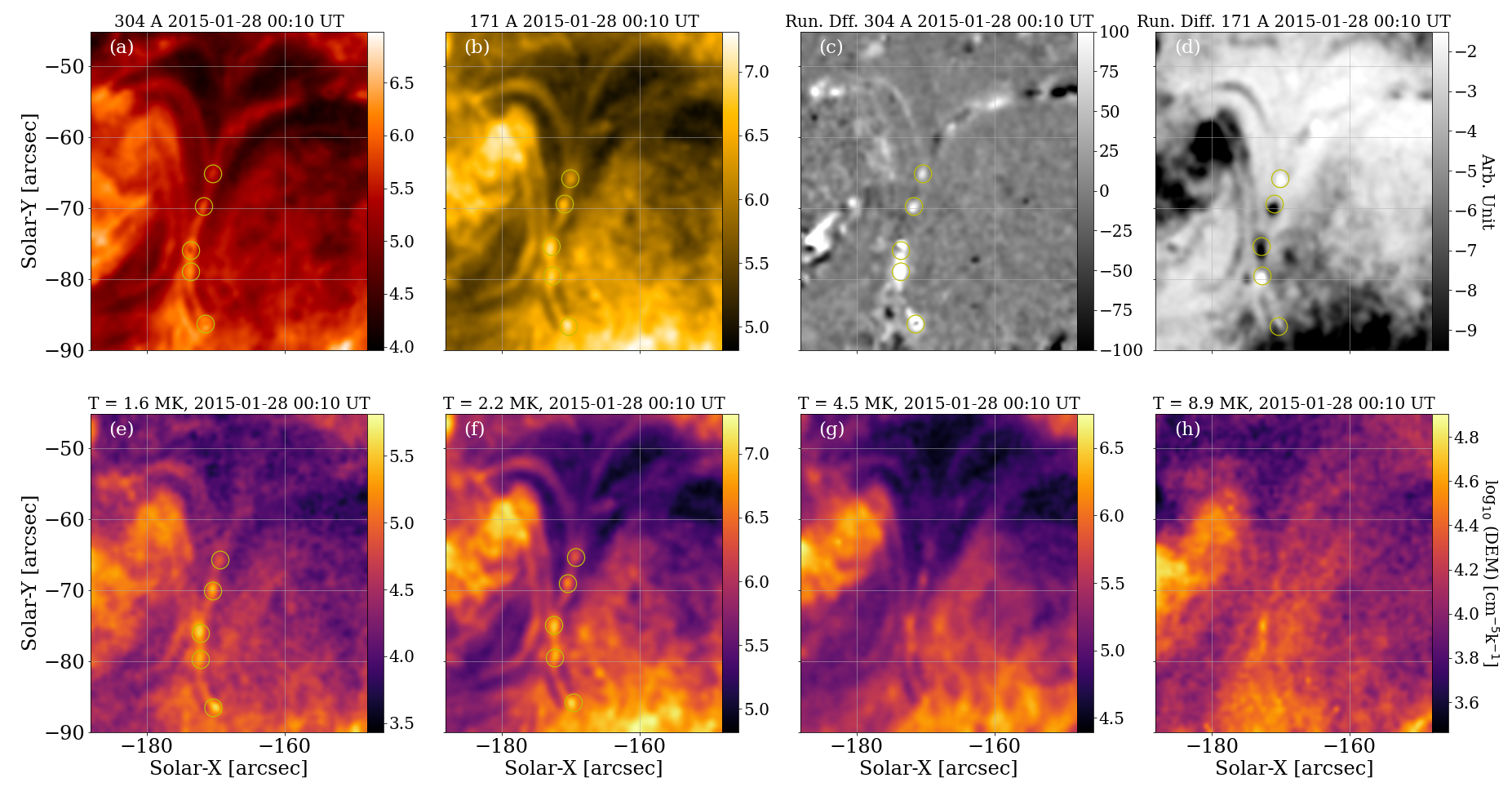}
\caption{Presence of plasmoid features at the interaction region in between the ambient EUV loops and filament channel in intensity and running-difference images along with associated DEM estimates. Panels (a) and (b) show \textit{SDO}/AIA 304, 171 $\textup\AA$ images at 00:10 UT on 28 January, with observed plasmoid features highlighted in circle. Corresponding running-difference images ((c), (d)) show these features as relatively bright/dark structures, indicating possible POS motions. Panels (e)-(h) suggests multi-thermal components of plasmoid structures with these features prominently visible in 1.6 MK (e) and 2.2 MK (f) temperatures. }\label{fig:6}
\end{center}
\end{figure*}

\begin{figure*}
\begin{center}
\includegraphics[scale=0.34]{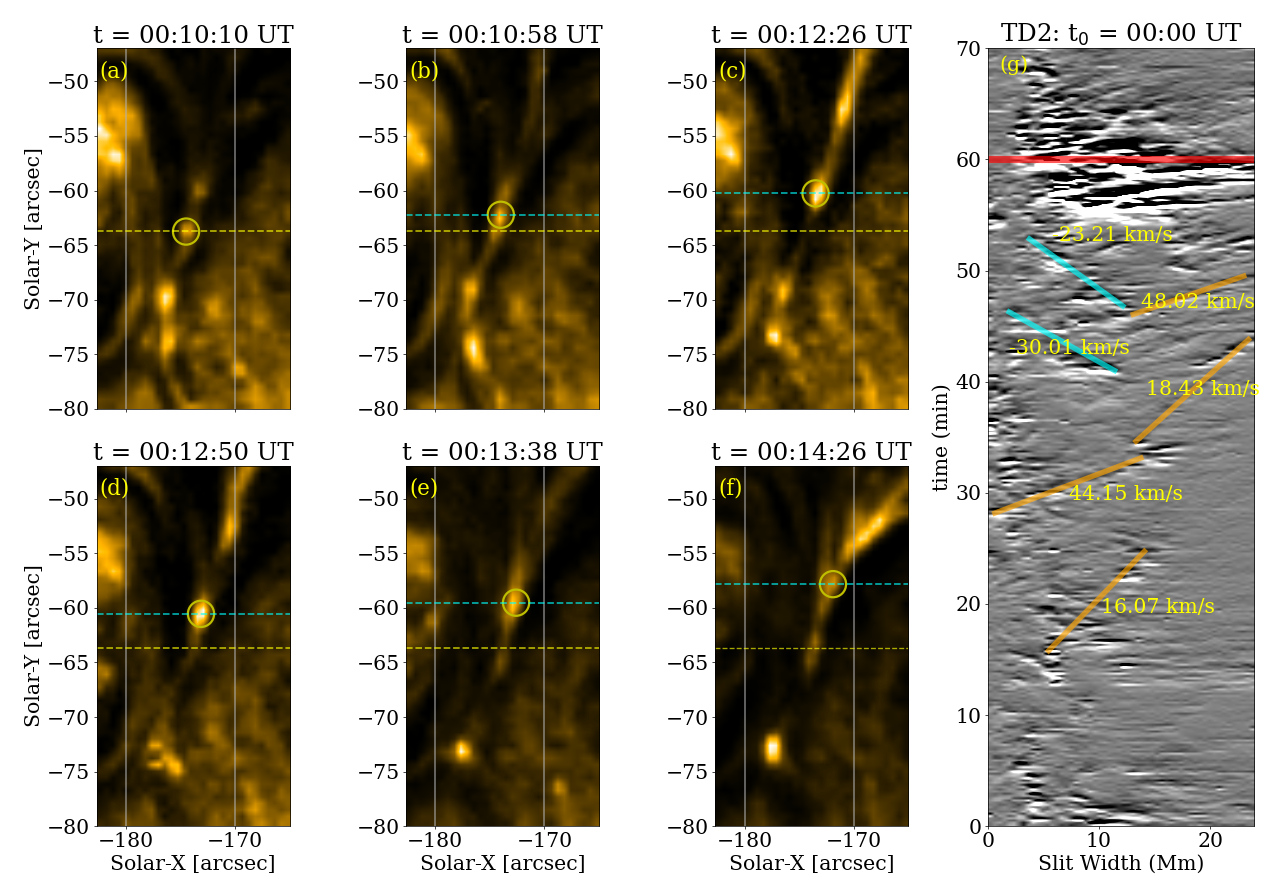}
\caption{Kinematics of the plasmoid features suggesting plasma (out)flows at the interaction region. Panels (a) - (h) show an example of projected POS motion of a plasma blob (circled) in \textit{SDO}/AIA 171 $\textup\AA$ images, during 00:10:10 - 00:14:26 UT. Initial position of the feature is marked by a dashed horizontal line (yellow) at 00:10:10 UT (a), with successive/relative upward motion highlighted by dashed lines (cyan) in consecutive images. Time-distance plot (g) from an artificial slit at the interaction region (TD2), suggests both upflow and downflow motions, with associated velocities. Horizontal line (red) marks the time for brightening (BR3, Figs. \ref{fig:2}(c1) - (c3)). An animation of this figure is available online.}\label{fig:7}
\end{center}
\end{figure*}

\begin{figure*}
\begin{center}
\includegraphics[scale=0.35]{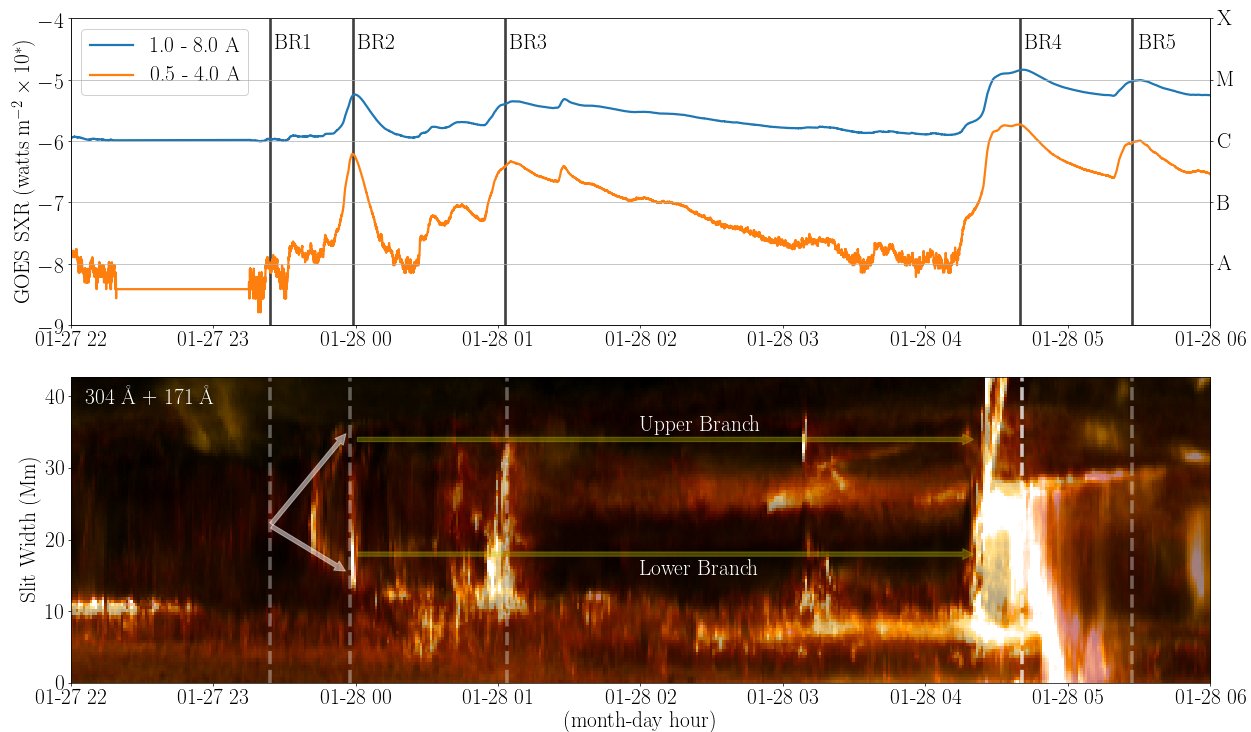}
\caption{Panels show the temporal evolution of AR12268  energetics, along with the time-distance plot of the bifurcated filament MFR. Top: \textit{GOES} 1.0-8.0 $\textup\AA$ and 0.5-4.0 $\textup\AA$ flux suggests two C-class flares, associated with brightening (BR2, BR3) in \textit{SDO}/AIA imaging data, following the photospheric reconnection/cancellation event (BR1) at S1 (Fig.\ref{fig:1}(b)), along with possible forking in filament structure after $\sim$23:48 UT on January 27, 2015. The M1.4 flare (BR4) erupts at 04:30 UT, followed by another C-class flare (BR5). Similar characteristics are also evident from the composite time-distance plot (bottom panel), and are marked by vertical dashed lines, along with `double-decker' configuration of the filament channel. The bifurcated filament channel with upper (UB) and lower branches (LB) are highlighted by horizontal lines. An animation of this figure is available online.}\label{fig:8}
\end{center}
\end{figure*}

\begin{figure*}
\begin{center}
\includegraphics[scale=0.225]{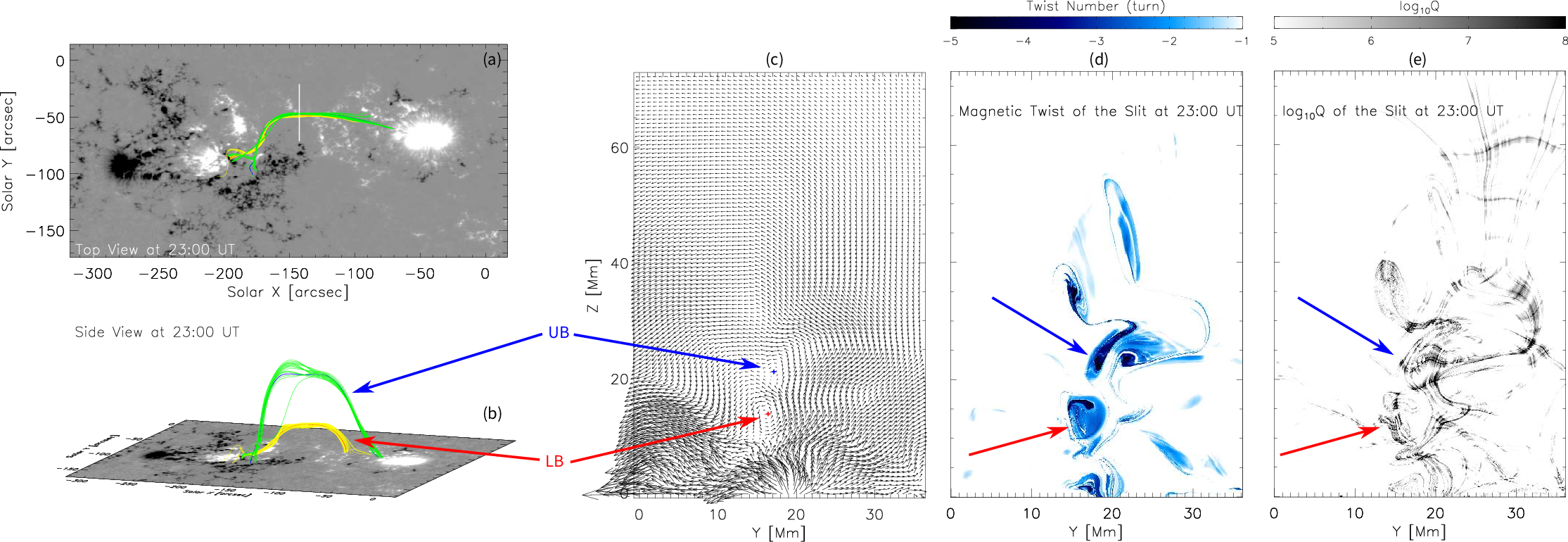}
\caption{Panels show extrapolated 3D NLFFF configuration corresponding to the pre-flare topology of the filament MFR at 23:00 UT on 27 Jan, located in between the active regions (AR12268 and AR12270). Panel (a) show the line-of-sight view of the active region, with selected field lines at the location of an artificial slit (marked by a vertical line). Panel (b) show the corresponding side-view with upper branch (UB) and lower branch (LB) substructures, along with the vertical cut of the vector magnetic field (c), perpendicular to the flux rope axis in yz-plane. The two rotating regions indicate presence of two flux ropes, very close to each other, with corresponding high twist (d) and squashing factor $Q$ (e). Field lines shown in panels (a) and (b) are traced from these regions, marked in panels (c) by arrows.}\label{fig:9}
\end{center}
\end{figure*}

\begin{figure*}
\begin{center}
\includegraphics[scale=0.225]{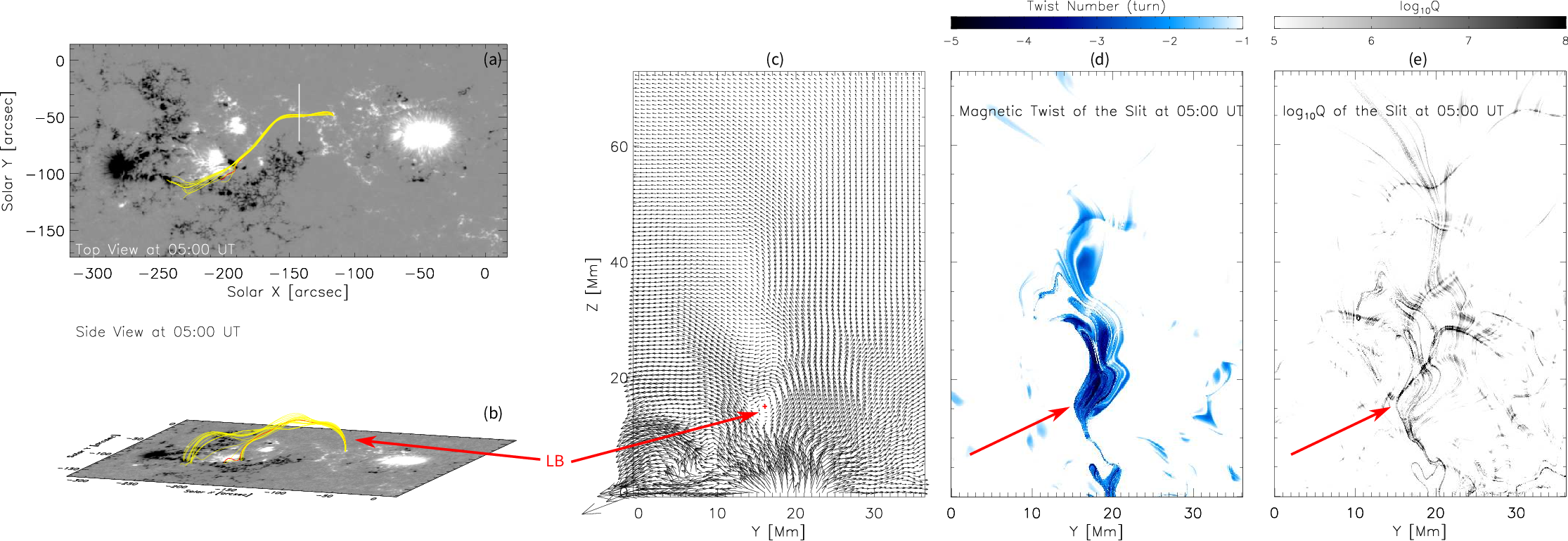}
\caption{Similar to Fig.\ref{fig:9}, but at 05:36 UT on 28 Jan, highlighting the post-flare magnetic topology and estimated twist (T$_{W}$) and squashing parameters ($Q$). Panel (d) and (e) indicates regions with high twist and $Q$ magnitudes, but without any apparent evidence of the upper branch (UB). The lower branch (LB) still remains at the height of 17 Mm in solar atmosphere with a T$_{W}$ of -1.60.}\label{fig:10}
\end{center}
\end{figure*}

\section{Results}

NOAA AR12268 first appeared at the eastern-limb of the solar-disk on January 23, 2015, with an initial magnetic configuration $\beta$. As the sunspot traversed through the visible disk, its magnetic complexity evolved from $\beta$ to $\beta\gamma$. Over time, eventually it generated around 20 flares (5 M- and 15 C-class). On January 28, 2015, the region was located near the disk-center (Fig. \ref{fig:1}(a)), at S11$^{\circ}$E10$^{\circ}$ and harbored a small filament MFR, identified in the chromospheric wavelengths (H$\alpha$ and \ion{He}{II}), observed from both space-borne and ground-based instruments (\textit{SDO}/AIA, ARIES-H$\alpha$). The filament structure (Fig. \ref{fig:1}(b)) was located between two active regions, with its lower leg (FP1) anchored in negative magnetic polarity (near AR12268), while an other leg (FP2) being rooted in the positive magnetic polarity located north-west, in close proximity to AR12270. This part (north-west) of the filament channel was significantly less active as compared to the other end (located near the AR12268), possibly due to higher flux concentrations but less emergence/cancellations of the magnetic field.

\subsection{Precursor brightening(s) and jet-like feature}
Examination of the imaging data from \textit{SDO}/AIA and HMI magnetograms for AR12268 revealed a highly complex environment, with frequent brightening from $\sim$22:47 UT on January 27, near the lower leg (FP1) of the filament MFR. Line-of-sight (LOS) magnetograms from \textit{SDO}/HMI (Fig. \ref{fig:1}(b): top panel) suggest bipolar magnetic flux in the region at sites associated with the corresponding brightening in the EUV channels. At $\sim$23:26 UT, brightening (BR1) was observed at site (S1), as enhanced emission in the AIA passbands associated with a jet-like feature, adjacent to FP1 of the filament structure. Despite of the strong background emissions, the concomitant dynamics of the flux feature were clearly identified in both, intensity and running-difference images.

The small localized brightening (BR1) at $\sim$23:26 UT (Figs. 2(a1)-(a3)), initiated the jet-like activity at $\sim$23:42 UT, visible in SDO/AIA 1600, 304 and 171 $\textup\AA$ channels. The structure appeared to move upwards, as suggested by the relative northeast shift with respect to the dark filament structure. The upward motion extended the brightening to nearby closed topology features at site (S2), observed at $\sim$00:00 UT as BR2 (Figs. \ref{fig:2}(b1)-(b3)), north to the filament leg (FP1). Figures \ref{fig:3}(a)-(c), show the evolution of the feature in intensity and running-difference images along with a time-distance plot highlight the swinging motion of the jet-like feature (Fig. \ref{fig:3}(d)). The feature with thread-like structures also showed displacement from the mean axis akin to a flux tube with confined kink wave.

\subsection{Interaction(s) between the ambient EUV loops and filament channel}

Subsequent to the brightening (BR1) at site S1, distinct EUV loop-like structures propagated towards the filament MFR (Figs. \ref{fig:5}(a)-(c)). These EUV loops interacted with the filament at around 00:00 UT on 28 January and resulted in a C-class flare. The shift of EUV loops appeared greater towards the northwestward direction (Figs. \ref{fig:5}(c), A2), as compared to the displacement towards the filament channel (A1), possibly due to the squashing between EUV loops and the filament channel at the interaction site. The accumulation of these EUV loop structures near the eastern section of the filament formed a V-shaped cusp structure (Fig. \ref{fig:4}). The kinematics of the EUV loops is also studied using an artificial slit (TD1), as shown in Figure \ref{fig:4}. Time-distance plot (TD1, Fig. \ref{fig:5}(d)) shows continues motion towards the filament feature during 23:42 - 00:00 UT on 27-28 January, with an average POS projected velocity of 4.18 km/s. It must however, be noted that the epoch of the flux migration at $\sim$23:42 UT concur with the observation of jet-like feature adjacent to the filament structure.

 In the highly complex magnetic environment of AR12268, the V-shaped cusp feature (Fig. \ref{fig:4}) formed due to the dynamics of post-reconnection magnetic fields highlights the role of magnetic topologies in sustaining this feature. It must be noted that any displacement of the magnetic field lines in a low-$\beta$ atmosphere is restored by the action of magnetic tension force. However, in the observed scenario, the feature remains in shape for a certain duration before diminishing from the observed passbands suggesting the complex magnetic structuring in the region. No significant photospheric shearing motions were observed at the site of cusp formation. Furthermore, the interaction between the coalesced EUV loops and the filament channel resulted into two oppositely-directed arc-shaped features, resembling X-type null-point topology, typical of complex/multiple source active-regions. At the interaction region between the ambient EUV loops and filament channel, bright plasmoid features (Fig. \ref{fig:6}) were also observed. Presence of these features indicates the possibility of high-density current sheets and associated instabilities at the reconnection site \citep{Takasao2012, Li2016, Gou2019}.

Figure \ref{fig:6} show the plasmoid features in co-temporal intensity, running-difference and emission measure maps at dominant temperatures. The coalesced EUV loops in close proximity to the filament feature are distinctly visible in DEM maps at 1.6 MK and 2.2 MK temperatures (Figs. \ref{fig:6}(e), (f)), as compared to the background, unperturbed corona. Bright plasmoids, as chain of plasma blobs, are also prominent at these temperatures indicating a multi-thermal plasma distribution within these structures. At higher temperatures (4.5 - 8.9 MK, Figs. \ref{fig:6}(g),(h)), the plasmoids tend to diffuse with the emissions from the interaction region.

A quantitative indication of the reconnection process between the ambient/coalesced EUV loops and the filament channel comes from the bi-directional plasma (out)flows at the interaction region. The magnitude of the POS projected velocities was estimated using an artificial slit (TD2, Fig. \ref{fig:4}) over \textit{SDO}/AIA 171 $\textup \AA$ images. Figures \ref{fig:7}(a)-(f) show an example of an upward ejection of a plasma blob at the interaction site, traced in unsharp-masked 171 $\textup \AA$ images. These (blob) motions tend to move towards the filament structure in the northwestwards direction, along the current sheet in between the EUV loops and the filament. The projected flow velocities (V$_{out}$) of these features (Fig. \ref{fig:7}(g)), during 00:00 - 01:00 UT on January 27, were found in the range 14.15 - 48.02 km/s, consistent with other reported cases \citep{Yokoyama2001a, Xue2016}. However, it must be noted that upward flows were dominant in the interaction region as compared to the downward motions. This intermittency could possibly be due to the projection effects, shift in X-shaped region, and/or inhomogeneity of the flux inflows. Furthermore, the reconnection rate is estimated in terms of Alfv\'en Mach number (M$_{A}$) of inflow velocity, with the assumption that outflow velocities reach the Alfv\'en velocities in the solar atmosphere. Taking the coalescing EUV loop velocity as inflow velocity (V$_{in}$), the estimated M$_{A}$ $\approx$ V$_{in}$/V$_{out}$ magnitudes were in the range of 0.29 - 0.08, consistent with earlier studies \citep{Yokoyama2001a, Takasao2012, Su2013, Xue2016}.

\subsection{Filament bifurcation}

The successive interactions between the filament with ambient loops resulted in enhanced EUV emissions from the region, followed by another brightening/reconnection event (BR3, Figs. \ref{fig:2}(c1)-(c3)) at 01:00 UT on 28 January. This event was measured as a C-class flare in \textit{GOES} soft x-ray (SXR) flux (Fig. \ref{fig:8}), and accompanied destabilization of magnetic flux in the region, as evident from the time-distance plot (TD2: Fig. \ref{fig:7}(g)). As a consequence, the filament channel showed clear indications of vertical splitting in its structure, observed in AIA intensity images with the filament MFR forked into two branches (Fig. \ref{fig:2}(d2)), forming a ``double-decker" configuration.
 
To better understand the evolution of the splitting behavior in the filament MFR and its possible association with the AR12268 energetics, a time-distance plot (marked as TD3, Fig. \ref{fig:4}) is constructed by placing an artificial slit over the feature. Figure \ref{fig:8} shows the TD plot, along with some SXR emission as seen in the \textit{GOES} (1.0 - 8.0 $\textup \AA$ and 0.5 - 4.0 $\textup \AA$) light curves. The enhancements in the SXR emission coincide with that in EUV intensity (304 $\textup\AA$ and 171 $\textup \AA$) for the split structure. The separation between both branched substructures, observed as dark features in TD plot, becomes prominent after the flux cancellation/reconnection events. The rapid splitting in the filament MFR coincide with the observations of jet-like flux feature at $\sim$23:47 UT (Figs. \ref{fig:3}(b)), associated with BR1. However, it must be noted that, the separation between the two branches appears nearly constant after 00:00 UT (BR2) in TD plot, which could possibly be due to the overlying magnetic field and/or LOS projection effects.

\subsection{NLFFF modeling and MHD instabilites}
To understand the magnetic field evolution of the bifurcated filament structure, NLFFF extrapolation were used which are constrained by the LOS magnetic field observations from \textit{SDO}/HMI. Snapshots in Figure \ref{fig:9} show the pre-flare topology of the magnetic field, the estimated twist number of the filament MFR and the squashing factor ($Q$). Accurate identification of the branched filament substructures was done using the vector magnetic field in \textit{yz}-plane (Fig. \ref{fig:9}(c)) at the site of an artificial slit (Fig. \ref{fig:9}(a)), along with associated twist number (T$_{W}$) and squashing ($Q$) parameters. The vector magnetic field for B$_{y}$ and B$_{z}$ estimates indicate the presence of two rotating structures, upper (UB) and lower branch (LB), with the central axis marked as `+' (Fig. \ref{fig:9}(c)). This stack of two distinct flux ropes also has relatively high magnitudes of twist number and squashing factor (Figs. \ref{fig:9} (d), (e)). Field lines traced from these locations highlights the presence of \textit{at least} two branches (LB and UB) associated with the filament structure with LB located at the height of 17 Mm in solar atmosphere. These branched substructures appeared to be separated at height of $\sim$8 Mm, (Figs. \ref{fig:9}(a), (b)) at 23:00 UT on January 27, though the bifurcation becomes evident at 01:00 UT on January 28 in \textit{SDO}/AIA imaging data. 

\begin{figure}
\begin{center}
\includegraphics[scale=0.4]{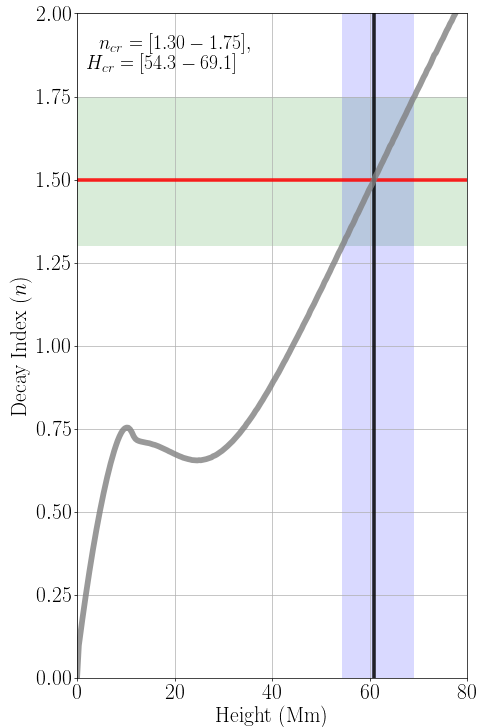}
\caption{Plot shows the variation in the estimated decay index ($n$) with height (H) over the photosphere. The theoretical range for the critical decay index ($n_{cr}$ = 1.3 - 1.75) is highlighted as shaded-region, along with the corresponding range ($H_{cr}$ = 54.3 - 69.1 Mm) over height. The specific value ($n$ = 1.5) for the onset for the torus instability is also marked with a horizontal line along with the height ($\sim$61.0 Mm), marked with a black vertical line.}\label{fig:11}
\end{center}
\end{figure}

The estimated twist number (Fig. \ref{fig:9}(c)) further suggests that the flux systems (LB and UB) had high twist initially, with -1.62 and -2.40 turns respectively. There was an apparent rise in the twisted UB of the filament structure, however, this rise is not prominent in time-distance analysis (Fig. \ref{fig:8}), possibly due to LOS projection effects. The relaxation of the unstable UB could result from the reconnection taking place between the filament MFR and the ambient EUV loops observed at the site of V-shaped cusp structure. The UB continues to rise up to $\sim$58 Mm in height that lies within the range (54.3-69.1 Mm) of critical height ($H_{cr})$, estimated for the critical range [1.3-1.75] for the decay index ($n_{cr}$). At this height (Fig. \ref{fig:11}) over the photosphere, the conditions become favorable for an ideal MHD instability (torus) to set-in that can further result into the loss of equilibrium of the flux system. The UB flux structure erupts at $\sim$04:30 UT as M1.4 flare (BR4; Figs. \ref{fig:2}(c1) - (c3), Fig. \ref{fig:8}), while the LB does not lift from the site throughout the process. This is further confirmed by NLFFF extrapolation at 05:36 UT on January 28 (Figs. \ref{fig:10}(a), (b)), with magnetic field vectors showing no presence of UB (Fig. \ref{fig:10}(c)).

The flare was also observed by the ground-based ARIES-H$\alpha$ telescope (Figs. \ref{fig:2}(a4) - (f4)), with clear post-flare ribbons. Figures \ref{fig:10}(c) - (e), suggests the complete removal of the UB, with slight change in T$_{W}$ magnitudes of |0.2| for LB substructure in pre- and post-flare estimates. Part of the erupted plasma appeared to follow the overlying flux arcade and then it drained back to the surface, in the active region AR12270. The region had continuous brightening which was mostly associated with AR12268 during the course of filament fragmentation to the partial eruption. \textit{GOES} emissions indicate another C-class flare (BR5) at $\sim$05:30 UT (Figs. \ref{fig:2}(f1)-(f3)) associated with a small-scale brightening in AR12268, however, this had no influence on the remaining low-lying LB structure. \textit{SoHO}/LASCO/C2 observed a weak CME event with structure leaving south-east quadrant at 06:00 UT on January 28. However, a clear association in between the M1.4 flare and this CME cannot be made due to the plasma deficit erupting structure and unclear observations of the CME.

\section{Discussion}
Our analysis of multi-wavelength observational data for 27-28 January, 2015 event, provide a clear indication of a link between the small-scale, localized photospheric brightening event(s), with large-scale coronal flaring/mass-ejection activities. The process involved mechanisms that include destabilization of the existing magnetic topology, reconnection and associated kinematic variations, together with magnetohydrodynamic instabilities and eruptions. We investigated a region near the disk-center (Fig. \ref{fig:1}), that had a complex magnetic environment with two active regions (AR12268, AR12270). These ARs also hosted a filament structure that was observed as a dark (absorption) feature in most wavelengths used for the study.

The series of events that led to the M1.4 flare on 28 January, 2015 at 04:30 UT, commenced with a small brightening (BR1) at $\sim$23:26 UT on 27 January, near one of the footpoints (FP1) of the filament channel (Fig. \ref{fig:2}). This brightening was followed by a jet-like feature (Fig. \ref{fig:3}), that accompanied the destabilization of the magnetic topology in the region. The perturbed magnetic field (EUV loops) migrated towards the filament channel (Fig. \ref{fig:5}) and formed a V-shaped cusp structure, along with a second brightening observed at $\sim$00:00 UT, north of the filament footpoint (FP1). The accumulated magnetic flux effectively \textit{shredded} the filament magnetic field through the process of reconnection, as evident from the observations of distinct plasmoid features (Fig. \ref{fig:6}), enhanced EUV/ x-ray flux and associated (out)flows (Fig.\ref{fig:7}). Interestingly, the interaction between the ambient magnetic field and the filament structure also resulted in the bifurcation of the filament into two substructures (UB and LB), as suggested in the time-distance plot (Fig. \ref{fig:8}) after 01:00 UT on 28 January. However, the presence of decked flux channels is also indicated at 23:00 UT on 27 January, in NLFFF extrapolations (Fig. \ref{fig:9}).

Previous interpretations of similar splitting behavior in filament MFRs were given by, e.g., \citet{Gibson2006a, Gibson2006}, where they suggested the reconnection in a central, vertical current sheet in the filament channel. They highlighted that shearing motions can lead to the formation of the current sheet within the flux rope structure, that eventually forms the site of the reconnection. If the filament mass is located near this (reconnection) site, then the plasma can accelerate upwards and later fall back, along with the observed vertical splitting of the filament MFR. Recently, \citet{Cheng2018} reported the vertical splitting in two filament cases and concluded the role of internal reconnection within the studied features. The enhanced emissions in both EUV and x-ray from a location between the erupting and remaining segments were interpreted as the signature of an internal reconnection process. Similar conclusions were also reported by \citet{Awasthi2018} for an eruptive flux system associated with a complex ejecta. \citet{Cheng2018}, however, also reported strong unwrith in split branches and suggested the loss of equilibrium due to an MHD instability, though they were not able to conclusively establish the nature of the instability responsible for the observed behavior.

Unlike the above, in our case, the split in the feature is associated with the interaction of filament structure with the ambient magnetic field after photospheric brightening event. Also, it must be noticed that no significant shearing motion were observed at the filament footpoints in the \textit{SDO}/HMI magnetograms. The possible explanation for the observed behavior comes from the numerical studies on filament flux splitting reported by, e.g., \citet{Kliem2014}. According to these authors, the reconnection in between the surrounding magnetic field and the filament channel is accompanied by the transfer of flux from ambient field to filament and vice-versa. A filament in a stable `double-decker' configuration can acquire additional magnetic flux from the ambient field that can enhance the twist in the upper branched (UB) substructure. This higher twist can potentially destabilize the UB from its prior position, while, the same acts as a stabilizing agent for the lower branch with an added overlying field. The UB, with additional twist and flux, can experience an upward stretch that can cause it to lift to a higher location while increasing the separation in between the both branches. Subsequently, the upper branch reaches to a height where an ideal MHD instability (torus) sets in and results in its removal as a partial eruption.

Similar cases of filament destabilization initiated by a precursor brightening event(s) were reported in a number of studies \citep[e.g.,][]{Chen2000,Joshi2016a,Chintzoglou2017,Bamba2019,Cheng2018,Awasthi2019,Dacie2018}. The filament feature in these reported cases, showed signatures of destabilization as splitting in the structure and/or as oscillations. Moreover, some of these studies \citep{Zhang2014,Li2016, Xue2016, Chintzoglou2017, Cheng2018} also highlighted the interaction of filament channel with the ambient magnetic field structures (e.g. loops, fibrils), along with common observables, like, untwisting/rotating jet-like motions, cusp-shaped features, high-density current sheets, plasmoids, and bidirectional flows, followed by the eruption of the filament MFR. 

\section{Conclusions}
In this paper, the partial eruption of a filament MFR at about 04:30 UT on January 28, 2015, is investigated that was located in a complex active region (AR12268) environment. A peculiar feature of the analysed case is the splitting of the filament channel into two distinct substructures, due to brightening event(s) in nearby flux systems and subsequent reconnection with ambient magnetic field. The onset of the M1.4 flare with eruption of the UB substructure is possibly due to an ideal MHD (torus) instability. The main conclusions from the study are as follows:

 \begin{enumerate}

\item{Hours before the onset of the M1.4 flare, localized, small-scale flux reconnection/cancellation events in the AR12268 were observed, that led to the propagation of bright EUV loops towards the filament structure. The region enclosed between the filament and the coalesced loops became the reconnection site with high current density, V-shaped cusp, plasmoids and bidirectional flows.}

\item{The interaction between the filament and the ambient magnetic field (EUV loops) destabilized the filament's magnetic topology, resulting into bifurcated substructures (UB and LB), evident from the time-distance analysis and NLFFF extrapolations. Our study highlights the role of the reconnection process between ambient and filament magnetic field in inducing the vertical splitting of the filament structures. Similar results from other studies \citep[e.g.,][]{Li2016, Xue2016, Chintzoglou2017} further supports our case, in regard to the split/partial-eruptions of filament structures observed in a region with complex magnetic topology.}

\item{3D NLFFF extrapolations of filament magnetic field and other estimated parameters (twist number, squashing factor), suggests two distinct substructures with higher twist in UB structure. The unstable UB rose to a higher altitude ($\sim$58 Mm) over photosphere due to added flux from external magnetic field, however, LB remained stable at 17 Mm. The height attained by UB was favourable for the onset of an ideal MHD instability, interpreted as the torus instability, from the estimated decay index ($n$) parameter.}

\item{The loss of equilibrium due to torus instability resulted in an M1.4 flare, along with the eruption of the UB substructure. However, the association of the eruption with any CME in \textit{SoHO}/LASCO observations was not clear.}

\end{enumerate}

Our study provides a key insight into the role of small-scale brightening events in the significant modification of an existing coronal topology that can further affect the nature of solar eruptive events. Apart from other magnetic field observables (location, orientation, magnitude, helicity), changes at much smaller scales pose a serious challenge to space weather prediction studies. Eruption of single/multiple flux segments due to such events can lead to single/complex CMEs in interplanetary medium, that can then have important consequences to the evolution and the dynamics of the near-Earth geomagnetic environment. Moreover, the unwinding action of the substructures may possibly transfer energy to the heating/expansion of erupted segments which will be the focus of future studies. The upcoming Daniel K. Inouye Solar Telescope, with the highest spatial/temporal resolution yet, will provide another significant step forward in this regard.

\section{Acknowledgment}
We thank Pankaj Kumar and Christopher Nelson for valuable discussions. \textit{SDO} is a mission of NASA`s Living with a Star Program, and respective teams of AIA and HMI, along with \textit{SoHO} and \textit{GOES} are thanked for providing free access to the data used in this work. Help from the technical staff at ARIES is also acknowledged for taking H$\alpha$ observations. RS and CC acknowledges support by the Spanish Ministry of Economy and Competitiveness (MINECO) through project AYA2016-80881-P (including FEDER funds). RC acknowledges the support from SERB-DST project no. SERB/F/7455/ 2017-17. RE and JL are grateful to Science and Technology Facilities Council (STFC, grant numbers ST/M000826/1) for the support received. JL also acknowledges support from STFC under grant No. ST/P000304/1. This research has made use of SunPy, an open source and free community-developed solar data analysis package written in Python \citep{TheSunPyCommunity2020}. 

\section{Data Availability}
The data underlying this article were accessed from Heliophysics Coverage Registry (HCR: http://www.lmsal.com/get\_aia\_data), which provides cutout data service for SDO/AIA data. The derived data generated in this research will be shared on reasonable request to the corresponding author.


\bibliographystyle{mnras}
\bibliography{mongaetal_ref}
\bsp	
\label{lastpage}
\end{document}